# Spontaneous nucleation and growth of GaN nanowires: Fundamental role of crystal polarity

Sergio Fernández-Garrido,* Xiang Kong, Tobias Gotschke, Raffaella Calarco, Lutz Geelhaar, Achim Trampert, and Oliver Brandt

*Paul-Drude-Institut für Festkörperelektronik, Hausvogteiplatz 5–7, 10117 Berlin, Germany*

E-mail: garrido@pdi-berlin.de

**Abstract**

We experimentally investigate whether crystal polarity affects the growth of GaN nanowires in plasma-assisted molecular beam epitaxy and whether their formation has to be induced by defects. For this purpose, we prepare smooth and coherently strained AlN layers on 6H-SiC(0001) and SiC(000$\bar{1}$) substrates to ensure a well-defined polarity and an absence of structural and morphological defects. On N-polar AlN, a homogeneous and dense N-polar GaN nanowire array forms, evidencing that GaN nanowires form spontaneously in the absence of defects. On Al-polar AlN, we do not observe the formation of Ga-polar GaN NWs. Instead, sparse N-polar GaN nanowires grow embedded in a Ga-polar GaN layer. These N-polar GaN nanowires are shown to be accidental in that the necessary polarity inversion is induced by the formation of $Si_xN$. The present findings thus demonstrate that spontaneously formed GaN nanowires are irrevocably N-polar. Due to the strong impact of the polarity on the properties of GaN-based devices, these results are not only essential to understand the spontaneous formation of GaN nanowires but also of high technological relevance.

Keywords: nanorod, nanocolumn, semiconductor, polarity, defect, nucleation

---

*To whom correspondence should be addressed

In crystal growth, certain materials are known to exhibit a pronounced tendency to spontaneously form nanostructures for specific growth conditions. Prominent examples of such materials include the technologically important semiconductors ZnO, for which the disposition to form a variety of nanostructures prevails even in homoepitaxial growth,[1–4] and GaN, which forms dense nanowire (NW) arrays in molecular beam epitaxy (MBE) on various crystalline as well as amorphous substrates such as $Si_xO$ and $Si_xN$.[5–11] These GaN NW arrays have evolved into prime candidates for a future generation of inexpensive and highly efficient light emitters,[12–17] since the NWs' crystal quality is high, independent of the substrate used for growth.[18–20] Despite the potential of these structures and the fact that they have been grown by MBE since the late nineties,[5,6] many aspects of the physical mechanisms governing their nucleation and growth are not understood.

Empirically, it is well established that GaN NWs grow exclusively under N-rich conditions with the polar (0001) axis parallel to the growth direction.[10,11,16,19,21] There is no concensus, however, about the actual NW polarity,[22–28] nor is it clear whether the substrate polarity plays any role in the spontaneous growth of GaN NWs.[23,26,27,29,30] Moreover, it is unknown whether the nucleation of NWs necessitates morphological or structural defects or occurs even on flat and defect free surfaces.[26–29,31,32]

Note that the polarity of the NWs has major consequences. As in the case of planar GaN layers, the NW orientation (Ga- or N-polar) is expected to manifest itself in a variety of chemophysical properties. For example, it determines the direction of both spontaneous and piezoelectric polarization, influences the incorporation of dopants and impurities, and affects the formation of native point defects.[33–37] In addition, the Ga- and N-polar surfaces have different chemical properties,[38] and the latter seems to favor the incorporation of In in the active region of light emitters.[39,40]

Despite its importance, the polarity of GaN NWs has not been the subject of focused research until very recently, when several groups demonstrated that GaN NWs on Si(111) are in fact N-polar.[25,26] For other substrates, such as $Al_2O_3$(0001), AlN/$Al_2O_3$(0001), or AlN/Si(111), however, only a few studies are available or none at all.[22–24,27,28] Due to the non-polar nature of the substrates which principally allows for the growth of NWs in either polarity, the findings of all of these studies are necessarily inconclusive.

The possible role of defects has not been addressed at all. On Si(111), the NWs actually grow on an amorphous Si$_x$N interlayer.[7,9,19,31,41] Nevertheless, it is universally reported that the NWs exhibit a pronounced biaxial texture following the orientation of the Si(111) substrate.[11,31,42,43] This finding strongly suggest the existence of pinholes in the amorphous Si$_x$N interlayer, resulting in partial epitaxy and simultaneously inducing NW nucleation. GaN NWs are also grown on Si(111) substrates protected by an AlN buffer layer which, however, contains a high density of threading defects due to the large lattice mismatch between AlN and Si.[26,27] Once again, these threading defects or the resulting morphological irregularities could be the decisive factor responsible for NW nucleation.[22,23,26–28,44]

In this letter, we present a study devoted to answering the question about the influence of polarity and defects on the spontaneous formation of GaN NWs. To this end, GaN NWs were grown on Al- and N-polar AlN buffer layers prepared on SiC(0001) and SiC(000$\bar{1}$) substrates, respectively. In contrast to Si(111) or Al$_2$O$_3$, the low lattice-mismatch between AlN and SiC (1%) together with the polar nature of the SiC substrate facilitate the growth of coherently strained AlN films with a well defined polarity in a layer-by-layer mode. Note that it is well established in the literature that the growth of III-nitride compounds on SiC(0001) and SiC(000$\bar{1}$) substrates results in metal- and N-polar III-nitride films, respectively.[45–47] Thus, this approach allows us to investigate the influence of the polarity on the spontaneous formation of GaN NWs on otherwise equivalent templates, and to elucidate whether defects are required or not to grow GaN NWs. It is shown that GaN NWs do grow on these high-quality substrates and, unlike epitaxial films, they do not necessarily preserve the polarity of the underneath buffer layer, since they are found to be always N-polar. In other words, even if the substrate polarity is expected to induce the formation of Ga-polar GaN NWs, only N-polar GaN NWs grow. In addition, we also found that the substrate polarity strongly influences the nucleation, distribution, and morphological properties of GaN NWs. Therefore, the results presented in this work show that defects are not required to grow GaN NWs, and demonstrate that crystal polarity plays a major role for the spontaneous formation of GaN NWs.

All samples were grown in a DCA Instruments P600 MBE system equipped with two radio-frequency $N_2$ plasma sources for active N, and solid-source effusion cells for Al and Ga. The as-received on-axis 6H-SiC(0001) and SiC(000$\bar{1}$) substrates were chemically cleaned and before initiating growth, 40 monolayers (ML) of Ga were deposited at 550°C and flashed-off at a temperature of 800°C to remove residual contaminants from the surface.[48] After that, 1×3 and 1×1 reflection high-energy electron diffraction (RHEED) patterns were observed for the SiC(0001) and SiC(000$\bar{1}$) substrates, respectively. The 1×3 RHEED pattern corresponds to a ($\sqrt{3} \times \sqrt{3}$)R30° reconstruction caused by 1/3 ML of Si atoms adsorbed on the surface.[49] Then an AlN buffer layer was grown under Al-rich growth conditions[50] at 800°C. To avoid plastic relaxation, its nominal thickness was kept at 6 nm, i. e., much below the critical thickness for strain relaxation. Afterwards, the surface was exposed to active N to ensure that there was no Al left prior to GaN growth. During AlN growth, independent of substrate polarity, the RHEED pattern was streaky, indicating that the growth proceeds in a two-dimensional (2D) mode [cf. insets of Figure 1(a) and (b)]. Upon AlN growth termination, distinct RHEED patterns were observed for SiC(0001) and SiC(000$\bar{1}$) substrates (1×3 and 1×1, respectively) revealing different reconstructions depending on substrate polarity. The surface morphology of the AlN buffer layer was examined by atomic force microscopy (AFM). As shown in Figure 1, for both polarities the AlN buffer layer exhibits a similar morphology with large 2D islands and a root-mean square roughness below 1 nm. After AlN growth, GaN was grown for 150 min under N-rich conditions using impinging fluxes of $\Phi_{Ga}$ = 4.5 nm/min and $\Phi_N$ = 11 nm/min to promote the spontaneous growth of GaN NWs.[21] Several samples were prepared using different growth temperatures (815 and 825°C). The morphological and structural properties of GaN NWs were investigated by scanning electron microscopy (SEM) and transmission electron microscopy (TEM). To assess the polarity of the samples, they were also analysed by either convergent beam electron diffraction (CBED) or electron energy-loss spectroscopy (EELS). The CBED simulations were performed by JEMS.[51] Cross-sectional TEM specimens were prepared using the standard method of mechanical grinding and dimpling down to below 25 µm followed by Ar-ion milling. TEM images were acquired using a JEOL JEM 3010 microscope operating at 200 kV and equipped with a Gatan charge coupled device camera and a Gatan Enfina parallel electron energy-loss spectrometer system. A 6 nm spot size was used for CBED and EELS measurements.

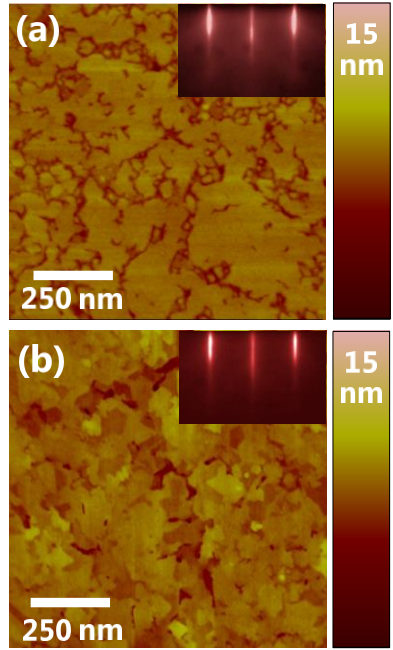

Figure 1: (Color online) AFM micrographs ($1\times1$ µm$^2$) of the AlN buffer layers grown on SiC(0001) (a) and SiC(000$\bar{1}$) (b) substrates. The insets show the RHEED pattern upon growth termination along the [11$\bar{2}$0] azimuth.

During the growth of GaN on the Al-polar AlN buffer layer, the RHEED pattern, consisting of inclined spots [cf. insets in Figure 2(a) and (c)], reveals a 3D growth mode. The resulting morphology for the sample grown at 815°C is shown in Figure 2(a) and (b), and is seen to be dominated by a faceted compact layer ($\approx$ 350 nm thick) interspersed with isolated GaN NWs aligned along the growth direction. This morphology is comparable to the usual one reported for GaN NWs grown on AlN/Al$_2$O$_3$(0001) and AlN/Si(111), where the AlN buffer layer frequently does not possess a single, well-defined polarity.[22,27,28] The NWs are $(1.2 \pm 0.1)$ µm long and exhibit high aspect ratios with diameters of $(21 \pm 6)$ nm. The density of NWs is approximately $6\times10^8$ cm$^{-2}$. This value is almost two orders of magnitude lower than the typical values reported for GaN NWs grown under similar conditions on bare Si(111).[8,11,21] Figure 2(c) and (d) illustrate how the morphology changes when the growth temperature is increased up to 825°C. As can be seen, the thickness of the faceted layer decreases (≈ 75 nm) and the NWs become shorter [(0.8 ± 0.2) µm] due to the exponential increase of both Ga desorption and GaN decomposition.[52] However, the NWs have very similar diameters [(22 ± 9) nm] and a comparable density ($5 \times 10^8$ cm$^{-2}$). These results contrast again with those reported for GaN NWs grown on bare Si(111), where higher growth temperatures lead to thinner and less closely spaced NWs.[11,21,53]

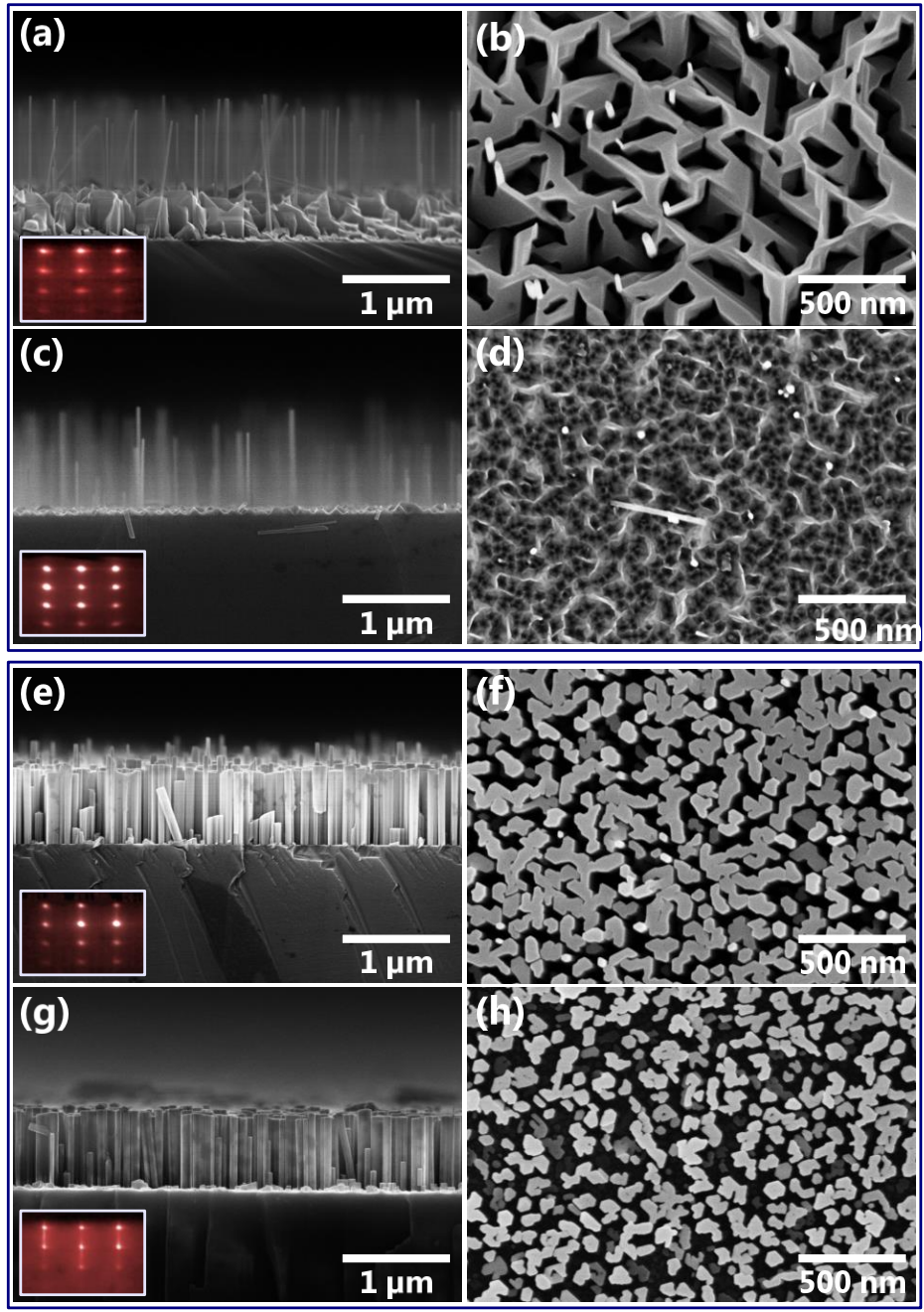

Figure 2: Cross-sectional (a)/(c) and plan-view (b)/(d) SEM images of GaN NWs grown on AlN-buffered SiC(0001) at 815/825°C. Cross-sectional (e)/(g) and plan-view (f)/(h) SEM images of GaN NWs grown on AlN-buffered SiC(000$\bar{1}$) at 815/825°C. The insets in figures (a), (c), (e), and (g) show the RHEED pattern along the [11$\bar{2}$0] azimuth upon growth termination.

Unlike for the growth on Al-polar AlN, the RHEED pattern observed during the growth of GaN on N-polar AlN coexhibits a superposition of sharp spots and dim streaks [cf. insets in Figure 2(e) and (g)], indicating an epitaxial relation as well as the presence of flat top facets. As shown in Figure 2(e)-(h), the morphology of the sample is indeed entirely different compared to the Al-polar one. No faceted compact layer is observed, but instead homogeneous and closely spaced GaN NWs with flat top facets (responsible for the dim streaks observed in the RHEED pattern). For the sample grown at 815°C, the NWs are (790 ± 80) nm long, have diameters of (60 ± 30) nm, and exhibit a density of $3 \times 10^{10}$ cm$^{-2}$ [Figure 2(e) and (f)]. In addition, as can be clearly observed in Figure 2(f), the NWs suffer from a high degree of coalescence. For increasing growth temperatures (825°C), the NWs become shorter [(730 ± 30) nm], the diameter decreases down to (50 ± 20) nm, and the density remains almost constant, $2.8 \times 10^{10}$ cm$^{-2}$. Consequently, due to the decrease in NW diameter, the sample grown at the highest temperature exhibits a lower coalescence degree [see Figure 2(f) and (h)], as typically observed for GaN NWs grown on bare Si(111).[11,21,53]

The SEM analysis of the samples presented above evidences that the mechanisms behind the spontaneous nucleation and growth of GaN NWs must depend on the polarity of the underneath AlN buffer layer since, for identical growth conditions, the morphology and distribution of the GaN NWs are found to be entirely different. In order to gain insight into the nucleation and growth mechanisms, the samples were further investigated by TEM, CBED, and EELS.

Figure 3 presents the TEM and CBED investigation of the samples grown on SiC(0001). High-resolution TEM (HRTEM) reveals that the Al-polar AlN buffer layer is continuous and, as shown in Figure 3(a), is 8 nm thick and coherently strained. Figure 3(b), (c) and (d) show, with different magnification, the GaN NWs and the faceted layer in between. The NWs grow either embedded into the faceted layer or along their sidewalls [Figure 3(c)], and they seem to nucleate on the AlN buffer layer [Figure 3(d)]. Their low density and the superimposed contrast from the faceted layer, however, makes it difficult to clarify this latter aspect conclusively. The analysis of the GaN faceted layer by CBED demonstrates that it is Ga-polar [Figure 3(b)] as expected for growth on an Al-polar AlN buffer layer. The lack of (0001) facets within the faceted layer is due to the fact that this facets are thermally highly unstable at the elevated temperatures used for the growth of GaN NWs. In fact, according to Ref. 52, at 815 °C the decomposition rate of a GaN(0001) film is as high as 5 nm/min, and this value is comparable to the impinging Ga flux. Since the isolated NWs were too thin to be investigated by CBED, their polarity was examined by EELS.

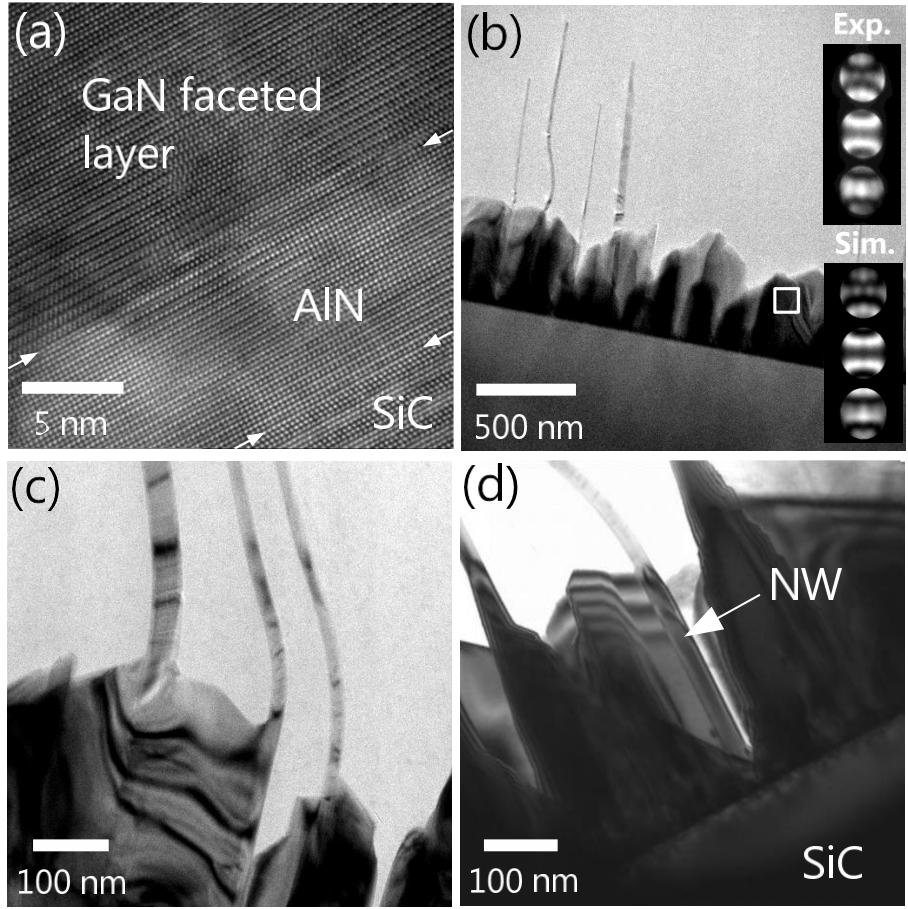

Figure 3: (a) Cross-sectional HRTEM image of the AlN buffer layer grown on SiC(0001) . The arrows indicate the SiC/AlN and AlN/GaN interfaces. (b-d) Cross-sectional TEM images of the GaN NWs grown on AlN-buffered SiC(0001) at 815°C. GaN NWs appear slightly bent due to sample preparation. The inset in figure (b) presents both the experimental and simulated CBED patterns of the GaN faceted layer in between GaN NWs for (0002), (0000), and (000$\bar{2}$) disks (from top to bottom) for a specimen thickness of 80 nm. The square indicates the region where the CBED patterns were acquired. Simulated CBED patterns for different specimen thicknesses are provided as supporting information.

EELS signals exhibit a remarkable difference between $(0002)$ and $(000\bar{2})$ Bragg conditions that can be used to identify NW polarity.[25,54] Figure 4(a) displays the EELS spectra of the GaN NWs grown on Al-polar AlN taken at $(0002)$ and $(000\bar{2})$ Bragg conditions after subtracting the background of the N K-edge and normalizing to the Ga L-edges. Based on the fact that the N K-edge peak at the (0002) orientation should show a higher intensity, the comparison with the TEM images [cf. inset of Figure 4(a)], which provide the growth direction, allow us to conclude that the NWs are N-polar, as shown in the inset of Figure 4(a). This result is consistent with the HRTEM analysis of the interface between the NWs and the faceted layer shown in Figure 4(b), which reveals the presence of an inversion domain boundary (IDB).

The inverted polarity exhibited by the GaN NWs grown on Al-polar AlN is a rather unusual finding because epitaxial GaN films grown on AlN-buffered SiC are known to preserve AlN polarity. Polarity inversion of $GaN(0001)$ to $GaN(000\bar{1})$ induced by heavy doping (particularly Mg) is, however, a widely observed phenomenon.[34,35,55] In our case, no Mg is present and the samples were not intentionally doped, but the Si adatoms forming the $(\sqrt{3} \times \sqrt{3})$R30° reconstruction of SiC(0001) constitute an abundant source of impurities. In fact, Rosa et al.[55] showed theoretically that a polarity flip of $GaN(0001)$ to $GaN(000\bar{1})$ may be induced if the GaN surface is exposed to Si under N-rich conditions. The polarity inversion may thus be explained as follows. During AlN growth, Si adatoms cannot be desorbed due to the extremely low Si vapor pressure, and Si incorporation into the crystal is hindered by the use of Al-rich growth conditions.[56] Consequently, the most probable scenario is that Si adatoms will segregate on the growth front until they become incorporated at the AlN/GaN interface, where the exposure of the AlN surface to active N and the subsequent growth of GaN under N-rich conditions favor Si incorporation. Assuming such a scenario, we propose the bonding configuration shown in Figure 5, which is based on the theoretical calculations reported in Ref. 55, to explain how Si may induce the formation of an IDB at the interface between the Al-polar AlN buffer layer and the N-polar GaN NWs. As illustrated in the figure, we suggest that the Si atoms segregated on the surface form a single Al-N bond with the Al-terminated AlN buffer layer, and three additional Si-N bonds that act as a seed for the growth of N-polar GaN NWs. The formation of N-polar NWs at this interface is natural for the growth conditions employed, but their appearance on a cation-polar substrate as such is an entirely accidental effect.

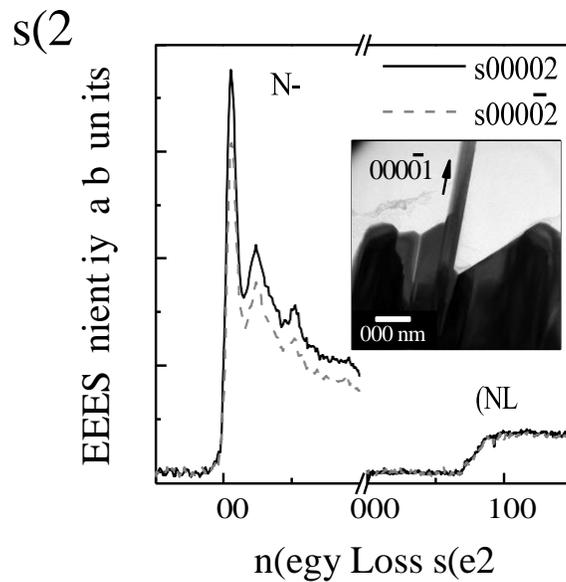

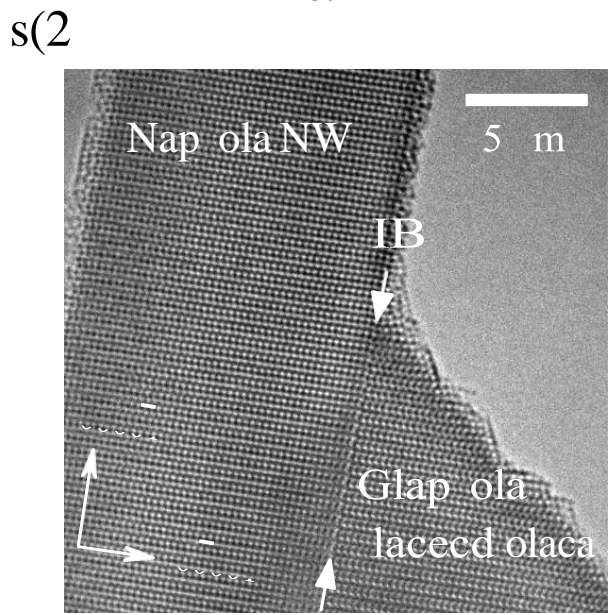

Figure 4: (a) A pair of EELS spectra of isolated GaN NWs grown on AlN-buffered SiC(0001). The spectra were acquired at (0002) and (000$\bar{2}$) Bragg conditions, and the N K-edge background was subtracted. The inset indicates the orientation of the GaN NW (N-polar). (b) HRTEM image of a GaN NW grown on AlN-buffered SiC(0001) at 815°C. The small arrows indicate the ID boundary at the interface between the NW and the faceted layer.

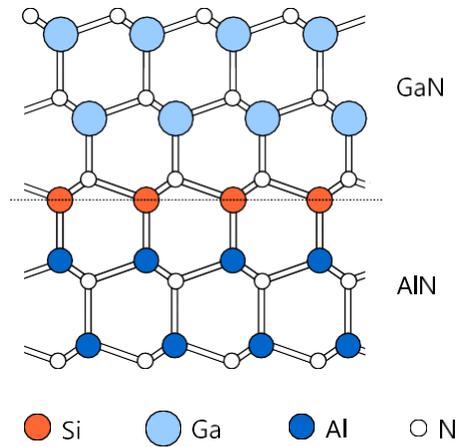

Figure 5: Schematic illustration of the Si-induced IDB at the interface between the Al-polar AlN buffer layer and the N-polar GaN NW.

Figure 6 shows the cross-sectional TEM images of the N-polar AlN buffer layer and the GaN NWs grown on SiC(000$\bar{1}$), together with the CBED measurements carried out to assess NW polarity (in this case the NWs were thick enough to be investigated by CBED). HRTEM images show that, as on SiC(0001), the AlN buffer layer is continuous, 8 nm thick and coherently strained. The GaN NWs are free of extended defects (except for those related to NW coalescence) and nucleate directly on the AlN buffer layer. In contrast to the growth on Al-polar AlN, we did not observe any trace of a compact GaN layer in between the GaN NWs. Regarding the NW polarity, the comparison of experimental and simulated CBED patterns [inset Figure 6(b)] demonstrates that they are N-polar. The latter result was also confirmed by EELS measurements (data not shown). In this case, the GaN NWs thus preserve the polarity of the AlN buffer layer.

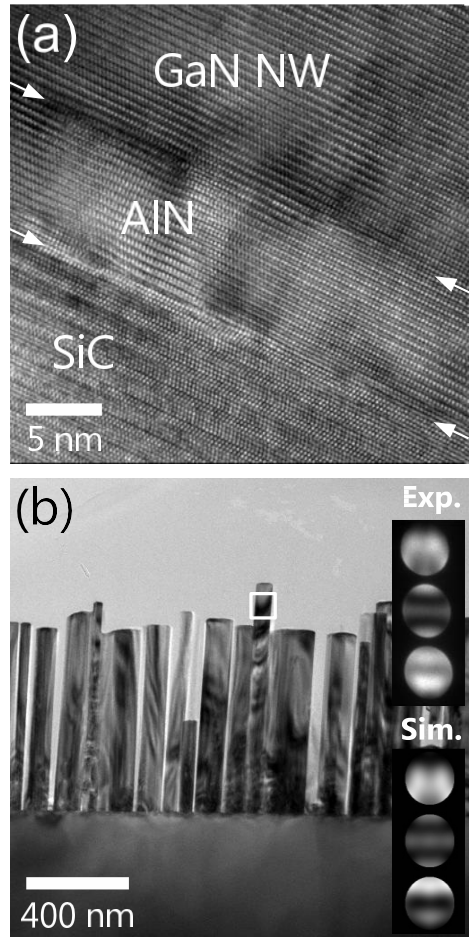

Figure 6: (a) Cross-sectional HRTEM image of the AlN buffer layer grown on SiC($000\bar{1}$). The arrows indicate the SiC/AlN and AlN/GaN interfaces. (b) Cross-sectional TEM image of the GaN NWs grown on AlN-buffered SiC($000\bar{1}$) at 815°C. The inset shows both the experimental and simulated CBED patterns for ($0002$), ($0000$), and ($000\bar{2}$) disks (from top to bottom) for a specimen thickness of 40 nm. The square indicates the region where the CBED patterns were acquired. Simulated CBED patterns for different specimen thicknesses are provided as supporting information.

To conclude, the central results of the present work are (i) the growth of dense and homogeneous N-polar GaN NW arrays on smooth and coherently strained N-polar AlN buffer layers, and (ii) the absence of any Ga-polar NWs even on Al-polar AlN buffer layers, which would be expected to define the polarity unequivocally. The former result reveals that the formation of GaN NWs does not require morphological or structural defects of the substrate, and is thus governed by spontaneous nucleation, in marked contrast to other materials.[44] The latter result points toward a fundamental difference between the (0001) and (000$\bar{1}$) surfaces for the spontaneous growth of GaN NWs. In fact, the few NWs observed on the Al-polar AlN buffer layer were N-polar, and the preceding polarity inversion is induced by impurities. In view of the present results, the spontaneous growth of Ga-polar GaN NWs on Al-polar AlN seems to be unlikely unless nucleation and growth is defect-induced, as reported, for example, in Refs. 22,23,44.

These results seemingly contrast with those recently published by Chen et al.[57], who demonstrated the growth of defect-free Ga-polar GaN NWs by metal organic vapor phase epitaxy (MOVPE). The possibility to grow Ga-polar GaN NWs by MOVPE but not by MBE can be the result of the inherent differences between these two growth techniques. MBE growth of GaN NWs is mainly governed by kinetic processes whereas the MOVPE growth of Ga-polar GaN NWs is carried out under very specific growth conditions that favor reaction-limited mechanisms rather than diffusion-limited (kinetic) transport.[57] In fact, the nucleation mechanisms underlying the MOVPE growth of Ga-polar GaN NWs seem to be completely different since the NWs exhibit a much lower density ($\approx 1 \times 10^3$ cm$^{-2}$), nucleate on pyramidal seeds, and have a pencil shape composed of a stem with vertical sidewall facets and a pyramidal top terminated by inclined semi-polar facets.

For MBE growth, a well known difference between the (0001) and (000$\bar{1}$) surfaces of GaN is that the kinetic barriers for the diffusivity of Ga and N adatoms, which depend on the III/V ratio (they are higher for III/V < 1), are lower for the former surface.[58] The higher diffusivity on the (0001) surface favors the lateral growth of GaN islands, and allows for the growth of GaN layers in a 2D mode under N-rich conditions for sufficiently high temperatures.[59] This fact can easily explain why we obtain a Ga-polar compact layer on Al-polar AlN instead of Ga-polar GaN NWs.

Another potential explanation lies in the fact that (0001) and (000$\bar{1}$) surfaces are expected to exhibit different surface energies. If, as it has been recently proposed by Dubrovskii *et al.*,[32] the dominant driving force for the spontaneous formation of GaN NWs is the anisotropy of surface energies, the present results could be eventually explained in terms of surface energetics. Unfortunately, the lack of reported values for the (0001) and (000$\bar{1}$) surface energies does not allow us to explore this possibility at present.

In this context, it is intriguing to compare our findings to those obtained for ZnO, a material also exhibiting a strong disposition for the spontaneous formation of NWs. Despite the obvious similarities between these materials with regard to crystal structure and related properties, ZnO NWs have recently be shown to be Zn-polar regardless of substrate polarity,[4] i.e, they are always cation-polar in contrast to the anion polarity observed for GaN NWs. In either case, NWs thus only form with a unique (but opposite) polarity.

Independent of the reason for the fundamental role of the polarity for NW formation, the present findings are essential for establishing an understanding of the spontaneous growth of GaN NWs, and of high technological relevance due to the major influence of the polarity on the properties of GaN-based electronic and optoelectronic devices.

## Acknowledgement

The authors would like to thank Bruno Daudin for stimulating discussions, Hans-Peter Schönherr and Claudia Herrmann for the maintenance of the MBE system, Anne-Kathrin Bluhm for the SEM images, Doreen Steffen for the preparation of TEM specimens, Sergey Sadofev for his help to calibrate the MBE system, and Uwe Jahn for SEM support and a critical reading of the manuscript.

## References

(1) Schmidt-Mende, L.; Macmanus-Driscoll, J. L. *Materials Today* **2007**, *10*, 40–48.

(2) Liu, J.; Xie, S.; Chen, Y.; Wang, X.; Cheng, H.; Liu, F.; Yang, J. *Nanoscale Res Lett.* **2011**, *6*, 619.

(3) Pfüller, C.; Brandt, O.; Flissikowski, T.; Grahn, H. T.; Ive, T.; Speck, J. S.; DenBaars, S. P. *Appl. Phys. Lett.* **2011**, *98*, 113113.

(4) Perillat-Merceroz, G.; Thierry, R.; Jouneau, P.-H.; Ferret, P.; Feuillet, G. *Nanotechnology* **2012**, *23*, 125702.

(5) Yoshizawa, M.; Kikuchi, A.; Mori, M.; Fujita, N.; Kishino, K. *Jpn. J. Appl. Phys.* **1997**, *36*, L459–L462.

(6) Sánchez-García, M.; Calleja, E.; Monroy, E.; Sánchez, F.; Calle, F.; Muñoz, E.; Beresford, R. *J. Cryst. Growth* **1998**, *183*, 23–30.

(7) Calleja, E.; Ristić, J.; Fernández-Garrido, S.; Cerutti, L.; Sánchez-García, M. A.; Grandal, J.; Trampert, A.; Jahn, U.; Sánchez, G.; Griol, A.; Sánchez, B. *Phys. Status Solidi B* **2007**, *244*, 2816–2837.

(8) Calarco, R.; Meijers, R. J.; Debnath, R. K.; Stoica, T.; Sutter, E.; Lüth, H. *Nano Lett.* **2007**, *7*, 2248–51.

(9) Stoica, T.; Sutter, E.; Meijers, R. J.; Debnath, R. K.; Calarco, R.; Lüth, H.; Grützmacher, D. *Small* **2008**, *4*, 751–4.

(10) Bertness, K. A.; Member, S.; Sanford, N. A.; Davydov, A. V. *IEEE J. Sel. Topics in Quantum Electron.* **2011**, *17*, 847–858.

(11) Geelhaar, L. et al. *IEEE J. Sel. Topics in Quantum Electron.* **2011**, *17*, 878–888.

(12) Mårtensson, T.; Svensson, C. P. T.; Wacaser, B. A.; Larsson, M. W.; Seifert, W.; Deppert, K.; Gustafsson, A.; Wallenberg, L. R.; Samuelson, L. *Nano Lett.* **2004**,

(13) Lagally, M. G.; Blick, R. H. *Nature* **2004**, *432*, 450.

(14) Kikuchi, A.; Kawai, M.; Tada, M.; Kishino, K. *Jpn. J. Appl. Phys.* **2004**, *43*, L1524–L1526.

(15) Sekiguchi, H.; Kishino, K.; Kikuchi, A. *Electronics Letters* **2008**, *44*, 151–152.

(16) Carnevale, S. D.; Yang, J.; Phillips, P. J.; Mills, M. J.; Myers, R. C. *Nano Lett.* **2011**, *11*, 866–71.

(17) Li, S.; Waag, A. *J. Appl. Phys.* **2012**, *111*, 071101.

(18) Calleja, E.; Sánchez-García, M.; Sánchez, F.; Calle, F. B.; Naranjo, F.; Muñoz, E.; Jahn, U.; Ploog, K. *Phys. Rev. B* **2000**, *62*, 16826–16834.

(19) Trampert, A.; Ristić, J.; Jahn, U.; Calleja, E.; Ploog, K. H. *Proceedings of the 13th International Conference on Microscopy of Semiconducting Materials, IOP Conf. Ser. No.* **2003**, *180*, 167.

(20) Brandt, O.; Pfüller, C.; Chèze, C.; Geelhaar, L.; Riechert, H. *Phys. Rev. B* **2010**, *81*, 45302.

(21) Fernández-Garrido, S.; Grandal, J.; Calleja, E.; Sánchez-García, M. A.; López-Romero, D. *J. Appl. Phys.* **2009**, *106*, 126102.

(22) Cherns, D.; Meshi, L.; Griffiths, I.; Khongphetsak, S.; Novikov, S. V.; Farley, N.; Campion, R. P.; Foxon, C. T. *Appl. Phys. Lett.* **2008**, *92*, 121902.

(23) Cherns, D.; Meshi, L.; Griffiths, I.; Khongphetsak, S.; Novikov, S. V.; Farley, N. R. S.; Campion, R. P.; Foxon, C. T. *Appl. Phys. Lett.* **2008**, *93*, 111911.

(24) Chèze, C.; Geelhaar, L.; Brandt, O.; Weber, W.; Riechert, H.; Münch, S.; Rothemund, R.; Reitzenstein, S.; Forchel, A.; Kehagias, T.; Komninou, P.; Dimitrakopulos, G.; Karakostas, T. *Nano Res.* **2010**, *3*, 528–536.

(25) Kong, X.; Ristić, J.; Sánchez-García, M.; Calleja, E.; Trampert, A. *Nanotechnology* **2011**, *22*, 415701.

(26) Hestroffer, K.; Bougerol, C.; Leclere, C.; Renevier, H.; Daudin, B. *Phys. Rev. B* **2011**, *84*, 245302.

(27) Brubaker, M. D.; Levin, I.; Davydov, A. V.; Rourke, D. M.; Sanford, N. A.; Bright, V. M.; Bertness, K. A. *J. Appl. Phys.* **2011**, *110*, 053506.

(28) Largeau, L.; Galopin, E.; Gogneau, N.; Travers, L.; Glas, F.; Jean-Christophe, H. *Cryst. Growth Des.* **2012**, *12*, 2724.

(29) Consonni, V.; Knelangen, M.; Geelhaar, L.; Trampert, A.; Riechert, H. *Phys. Rev. B* **2010**, *81*, 085310.

(30) Consonni, V.; Hanke, M.; Knelangen, M.; Geelhaar, L.; Trampert, A.; Riechert, H. *Phys. Rev. B* **2011**, *83*, 035310.

(31) Hestroffer, K.; Leclere, C.; Cantelli, V.; Bougerol, C.; Renevier, H.; Daudin, B. *Appl. Phys. Lett.* **2012**, *100*, 212107.

(32) Dubrovskii, V.; Consonni, V.; Trampert, A.; Geelhaar, L.; Riechert, H. *Phys. Rev. B* **2012**, *85*, 165317.

(33) Bernardini, F.; Fiorentini, V.; Vanderbilt, D. *Phys. Rev. B* **1997**, *56*, R10024–R10027.

(34) Green, D. S.; Haus, E.; Wu, F.; Chen, L.; Mishra, U. K.; Speck, J. S. *J. Vac. Sci. Technol., B* **2003**, *21*, 1804.

(35) Feduniewicz, A.; Skierbiszewski, C.; Siekacz, M.; Wasilewski, Z. R.; Sproule, I.; Grzanka, S.; R., J.; Borysiuk, J.; Kamler, G.; Litwin-Staszewska, E.; Czernecki, R.; Boćkowski, M.; Porowski, S. *J. Cryst. Growth* **2005**, *278*, 443–448.

(36) Chichibu, S. F.; Setoguchi, A.; Uedono, A.; Yoshimura, K.; Sumiya, M. *Appl. Phys. Lett.* **2001**, *78*, 28.

(37) Ptak, A. J.; Holbert, L. J.; Ting, L.; Swartz, C. H.; Moldovan, M.; Giles, N. C.; Myers, T. H.; Van Lierde, P.; Tian, C.; Hockett, R. A.; Mitha, S.; Wickenden, A. E.; Koleske, D. D.; Henry, R. L. *Appl. Phys. Lett.* **2001**, *79*, 2740.

(38) Seelmann-Eggebert, M.; Weyher, J. L.; Obloh, H.; Zimmermann, H.; Rar, A.; Porowski, S. *Appl. Phys. Lett.* **1997**, *71*, 2635.

(39) Xu, K.; Yoshikawa, A. *Appl. Phys. Lett.* **2003**, *83*, 251.

(40) Nath, D. N.; Gür, E.; Ringel, S. A.; Rajan, S. *Appl. Phys. Lett.* **2010**, *97*, 071903.

(41) Chèze, C.; Geelhaar, L.; Trampert, A.; Riechert, H. *Appl. Phys. Lett.* **2010**, *97*, 043101.

(42) Jenichen, B.; Brandt, O.; Pfüller, C.; Dogan, P.; Knelangen, M.; Trampert, A. *Nanotechnology* **2011**, *22*, 295714.

(43) Largeau, L.; Dheeraj, D. L.; Tchernycheva, M.; Cirlin, G. E.; Harmand, J. C. *Nanotechnology* **2008**, *19*, 155704.

(44) Bierman, M. J.; Lau, Y. K. A.; Kvit, A. V.; Schmitt, A. L.; Jin, S. *Science* **2008**, *320*, 1060–3.

(45) Stutzmann, M.; Ambacher, O.; Eickhoff, M.; Karrer, U.; Lima Pimenta, a.; Neuberger, R.; Schalwig, J.; Dimitrov, R.; Schuck, P.; Grober, R. *Phys. Status Solidi B* **2001**, *228*, 505–512.

(46) Wang, X.; Yoshikawa, A. *Prog. Cryst. Growth Charact. Mater.* **2004**, *48-49*, 42–103.

(47) Fossard, F.; Brault, J.; Gogneau, N.; Monroy, E.; Enjalbert, F.; Dang, L. S.; Bellet-Amalric, E.; Monnoye, S.; Mank, H.; Daudin, B. *Materials Science Forum* **2004**, *457-460*, 1577–1580.

(48) Brandt, O.; Muralidharan, R.; Waltereit, P.; Thamm, A.; Trampert, A.; von Kiedrowski, H.; Ploog, K. H. *Appl. Phys. Lett.* **1999**, *75*, 4019.

(49) Han, Y.; Aoyama, T.; Ichimiya, A.; Hisada, Y. *Surf. Sci.* **2001**, *493*, 238–245.

(50) Koblmüller, G.; Averbeck, R.; Geelhaar, L.; Riechert, H.; Hösler, W.; Pongratz, P. *J. Appl. Phys.* **2003**, *93*, 9591.

(51) Stadelmann, P. A. *Ultramicroscopy* **1987**, *21*, 131.

(52) Fernández-Garrido, S.; Koblmüller, G.; Calleja, E.; Speck, J. S. *J. Appl. Phys.* **2008**, *104*, 033541.

(53) Consonni, V.; Knelangen, M.; Trampert, A.; Geelhaar, L.; Riechert, H. *Appl. Phys. Lett.* **2011**, *98*, 071913.

(54) Kong, X.; Hu, G. Q.; Duan, X. F.; Lu, Y.; Liu, X. L. *Appl. Phys. Lett.* **2002**, *81*, 1990.

(55) Rosa, A.; Neugebauer, J. *Surf. Sci.* **2006**, *600*, 335–339.

(56) Hoke, W. E.; Torabi, A.; Mosca, J. J.; Hallock, R. B.; Kennedy, T. D. *J. Appl. Phys.* **2005**, *98*, 084510.

(57) Chen, X. J.; Gayral, B.; Sam-Giao, D.; Bougerol, C.; Durand, C.; Eymery, J. *Appl. Phys. Lett.* **2011**, *99*, 251910.

(58) Scheffler, M.; Zywietz, T. *Appl. Phys. Lett.* **1998**, *73*, 487–489.

(59) Koblmüller, G.; Fernández-Garrido, S.; Calleja, E.; Speck, J. S. *Appl. Phys. Lett.* **2007**, *91*, 161904.

**Table of Contents Graphic**

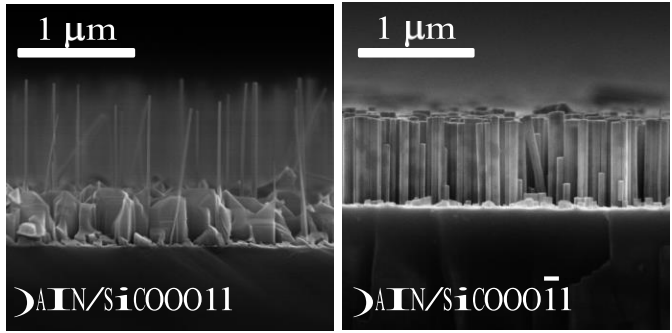